\documentclass{llncs}

\usepackage{amsmath}
\usepackage{graphicx}

\begin{document}

\title{Does the Red Queen reign in the kingdom of digital organisms?}
\titlerunning{Red Queen in digital organisms}  
%
\author{Claus O. Wilke}
\authorrunning{Claus O. Wilke}   
\institute{Digital Life Laboratory 136-93, California Institute of Technology\\Pasadena, CA 91125, USA\\
\email{wilke@caltech.edu}\\ WWW home page:
\texttt{http://dllab.caltech.edu/\homedir wilke}}

\maketitle              

\begin{abstract}
  In competition experiments between two RNA viruses of equal or almost equal
  fitness, often both strains gain in fitness before one eventually excludes
  the other. This observation has been linked to the Red Queen effect, which
  describes a situation in which organisms have to constantly adapt just to
  keep their status quo. I carried out experiments with digital organisms
  (self-replicating computer programs) in order to clarify how the competing
  strains' location in fitness space influences
  the Red-Queen effect. I found that gains in fitness during competition were
  prevalent for organisms that were taken from the base of a fitness peak, but
  absent or rare for organisms that were taken from the top of a peak or from
  a considerable distance away from the nearest peak. In the latter two cases,
  either neutral drift and loss of the fittest mutants or the waiting time to
  the first beneficial mutation were more important factors.  Moreover, I
  found that the Red-Queen dynamic in general led to faster exclusion than
  the other two mechanisms.
\end{abstract}

\section{Introduction}

Two major principles of evolutionary biology have been observed in competition
experiments between variants of RNA viruses with identical or almost identical
fitness: competitive exclusion and Red Queen
dynamic~\cite{Clarkeetal94,Queretal96}. The competitive exclusion principle
refers to the ultimate outcome of these competition experiments, and states
that when two or more species live on the same resource, eventually all but
one will die out~\cite{Hardin60}. The Red Queen dynamic refers to the initial
phase of these competition experiments, where the two competing virus variants
both increase in fitness while they remain in roughly equal concentrations.
Van Valen~\cite{vanValen73} had originally proposed the Red Queen dynamic as a
metaphor for the struggle for existence of species in complex ecosystems.
These species, just like the Red Queen in Lewis Carroll's \emph{Through the
  Looking Glass}, would  constantly have to run (that is, to adapt to changing
conditions) just to remain where they were (see also Ref.~\cite{Ridley94}).

Sol\'e et al.~\cite{Soleetal99} studied the competition of neutral virus
variants in a simple bitstring model with additive fitness, related to
Kauffman's $NK$ model~\cite{Kauffman90}, and could confirm both competitive
exclusion and Red Queen dynamic in
their model.  Moreover, Sol\'e et al. showed that the competitive exclusion
principle follows immediately from the quasispecies
equations~\cite{EigenSchuster79,Eigenetal88} that describe virus evolution.

The Red Queen effect, on the other hand, is not a necessary conclusion of the
quasispecies dynamic, as we can see from a simple thought experiment: Assume
that we allow two viruses to compete that are both situated on top of the
highest peak in the fitness landscape. None of the two viruses can accumulate
any further beneficial mutations, and the outcome of a competition between
them will be determined by genetic drift. Clearly, the Red Queen dynamic can
occur only if beneficial mutations are sufficiently abundant, so that
adaptation happens on a faster time scale than genetic drift. The additive
model of Sol\'e et al.\ has a fairly high rate of positive mutations for all
but the very best sequences in the fitness landscape, which explains why they
observed the Red Queen dynamic. Such simple additive or multiplicative fitness
landscapes lead to a smooth increase in average fitness over
time~\cite{Tsimringetal96,Rouzineetal2003}, and such increase has been
reported for RNA viruses~\cite{Novellaetal95}. However, in many cases
beneficial mutations are rare, which leads to the frequently-oberseved
stepwise increase in average
fitness~\cite{LenskiTravisano94,Elenaetal96,FontanaSchuster98,vanNimwegenetal99a}.

\begin{figure}[t]
\centerline{
\includegraphics[width=.8\columnwidth]{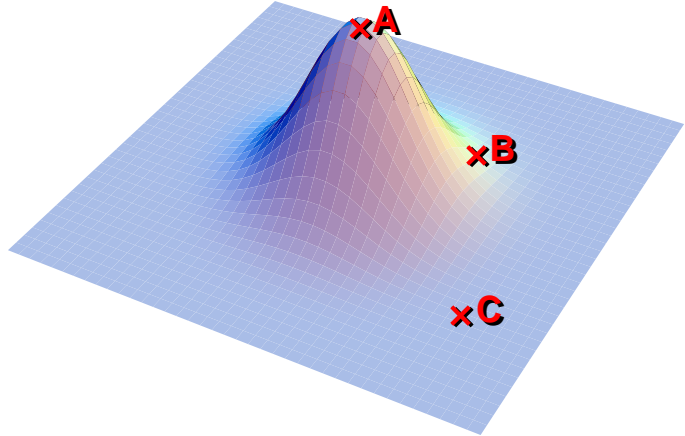}
}
\caption{\label{fig:fitness-landscape}Schematic illustration of the location
  in fitness space of the three strains used in this study. Strain A was taken
  from the top of a fitness peak, strain B from the base of a fitness peak,
  and strain C from some distance away from the nearest fitness peak. For
  simplicity of illustration, this figure shows only one fitness peak, whereas
  in reality the fitness peak of strain A was different from the one of
  strains B and C.
}
\end{figure}

Here, I present a study of the influence of the immediate neighborhood in the
fitness space on the competition dynamics of identical organisms. I carried
out this study with digital organisms (self-replicating computer
programs)~\cite{WilkeAdami2002}, using the Avida platform. Avida is similar to
Tom Ray's Tierra~\cite{Ray91}, and has been described extensively in the
literature~\cite{Adami98,Lenskietal99,Adamietal2000,Wilkeetal2001b}. In Avida,
the self-replicating computer programs are rewarded with increased CPU speed
when they perform certain logical computations. By changing the bonus
structure, that is, by changing which computations are rewarded with what
bonuses, the researcher can shape the fitness landscape in which the digital
organisms evolve. I studied three different strains of digital organisms that
were located on the top (A), at the base (B), and some distance away from the
nearest fitness peak (C) (Fig.~\ref{fig:fitness-landscape}).

\section{Materials and methods}

\subsection{Computer experiments}
I used Avida version 1.99, which is available from
\texttt{http://sourceforge.net/\\projects/avida}. I used the default setup,
apart from the following modifications: I switched off length changes, and set
insertion and deletion mutations to zero. As the original ancestor of all
organisms in this study, I used the handwritten organism
\texttt{organism.heads.100}, which comes with the Avida distribution. I used
two different environment files (the environment determines the logical
operations for which organisms are rewarded, and thus defines the fitness
landscape).  The first one, environment~1, did not reward any logical
computations, so that fitness was a direct function of an organism's
replication efficiency. The second one, environment~2, rewarded all possible
one-, two-, and three-input logical operations. In this environment,
replication efficiency was only a minor component of fitness, whereas the
successful completion of as many logical computations as possible resulted in
high fitness values. The mutation rate in all experiments was 0.75 miscopied
instructions per offspring organism, unless explicitly stated otherwise.

I created the strains A, B, and C (as illustrated in
Fig.~\ref{fig:fitness-landscape}) as follows. For strain A, I inocculated a
population of $N=5000$ organisms with the handwritten ancestor, and propagated
this population for 14,000 generations in environment~1. Then, I extracted the
most abundant organism, which replicated roughly two times faster than the
original ancestor. By generation 14,000, the population had not experienced
any significant fitness improvements for over 10,000 generations, which showed
that it had indeed reached the top of a fitness peak. For strain B, I first
evolved from the handwritten ancestor an organism that could perform all
possible one- and two-input logical computations. Then, I inocculated a
population of size $N=200$ with this evolved organism, in order to let the
organism accumulate mutations.  After 50 generations, I extracted from the
population a variant with drastically reduced (by a factor of fifteen)
fitness. This variant differed in six instructions (out of 100) from its
evolved ancestor, and had lost five logical operations. It was therefore at
the base of a fitness peak in environment~2, where many additional logical
computations could be learned. For strain C, I simply used the handwritten
ancestor (also in environment~2). The handwritten ancestor did not perform any
logical computations. In Avida, the evolution of the first logical computation
is harder than the evolution of additional ones, which guaranteed that C was
further away from the fitness peak than B.

I carried out all competition experiments with the following protocol. For
each strain, I used three different population sizes, $N=100$, $N=1000$, and
$N=10,000$. I did 100 replicates per population size and competing strain. In
each competition experiment, I inocculated the population with $N$ identical
copies of the strain under study, and marked half of them with an inheritable
neutral marker. Then, I propagated the population in the respective
environment (environment~1 for strain A, environment~2 for strains B and C)
at fixed size $N$ until the population consisted of either all marked or all
unmarked organisms. Each generation, I recorded the average and maximum
fitness of the marked and unmarked organisms, as well as their relative
frequencies.

\subsection{Coalescent theory}

For the competition of two strains in a completely flat landscape, we can
calculate the distribution of extinction times from coalescent theory. Let
$p_j(t)$ be the probability that after $t$ generations, each of the $N$
organisms in the population is a direct descendant of one of a group of only
$j$ organisms that were present at $t=0$. Further, let $m_j=[1-(1/2)^{j-1}]$
be the probability that this group of $j$ contained both marked and
unmarked organisms. Then, the probability that neither the marked nor the
unmarked strain is extinct after $t$ generations is
\begin{equation}\label{eq:extinct-neutral}
  P(t_{\rm extinct}>t) = \sum_{j=2}^N m_j p_j(t)\,.
\end{equation}
The quantity $p_j(t)$ has been given in various forms in the
literature~\cite{Griffiths80,Donnelly84,Tavare84}. The following version is
based on the result by Tavar\'e~\cite{Tavare84}:
\begin{equation}
  p_j(t)=\sum_{k=j}^N \rho_k(t)(-1)^{k-j}(2k-1)C_{jk}\,,
\end{equation}
where $\rho_k(t)=\exp[-k(k-1)t\sigma^2/(2N)]$, and $C_{jk}$ satisfies the
recursion (for $k\geq j$):
\begin{subequations}
\begin{align}
  C_{11}&=1\,,\\
  C_{jk+1}&=\frac{(N-k)(k+j-1)}{(N+k)(k-j+1)}C_{jk}\,,\\
  C_{j+1k}&=\frac{(k+j-1)(k-j)}{j(j+1)}C_{jk}\,.
\end{align}
\end{subequations}
The quantity $\sigma^2$ is the variance of the offspring distribution. In
Avida, it has the value $\sigma^2=2/3$, because organisms in Avida can have
either zero, one, or two offspring organisms per generation, with equal
probabilities (if all organisms have equal fitness).

\section{Results and Discussion}

\begin{figure}[b]
\centerline{
\includegraphics[width=.7\columnwidth]{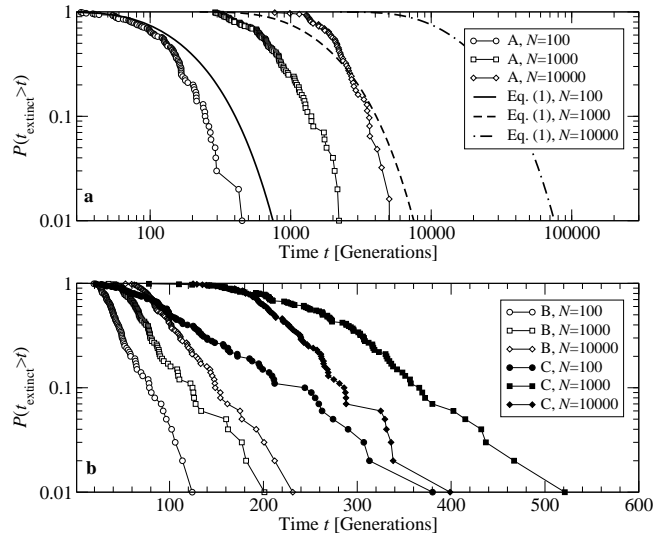}
}
\caption{\label{fig:extinct-times}Distributions of extinction times
  for all three strains and all three population sizes.  I plot the
  cummulative frequency of all extinction times larger than or equal to the
  given generation, $P(t_{\rm extinct}>t)$. Part a shows also the prediction
  for neutral evolution according to Eq.~\eqref{eq:extinct-neutral}.}
\end{figure}


I measured the time to extinction of the losing clone in all competition
experiments and derived the distributions of extinction times. The nine
distributions (three strains at three different population sizes) are shown in
Fig.~\ref{fig:extinct-times}. The extinction times are the longest for
competitions of strain A, intermediate for competitions of strain C, and
shortest for competitions of strain B. For strains A and B, the extinction
times grow systematically with increasing population size, whereas for strain
C, the extinction times grow from $N=100$ to $N=1000$, but then decrease again
for $N=10,000$. (Notice that the line with solid diamonds lies on the left of
the line with solid squares in Fig.~\ref{fig:extinct-times}b.)

\begin{figure}[b]
\centerline{
\includegraphics[width=.7\columnwidth]{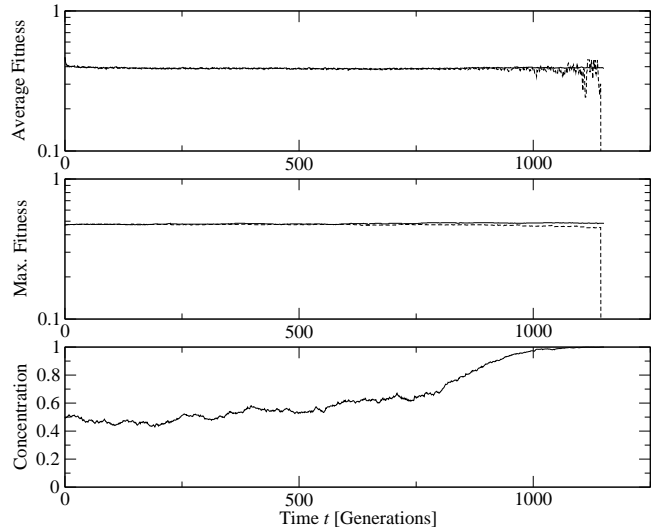}
}
\caption{\label{fig:compN10000-A}Typical dynamic of a competition experiment
  for strain A ($N=10,000$). The top two graphs show the average and the
  maximum fitness of the winning (solid lines) and the losing (dashed lines)
  clones as a function of time. The bottom graph shows the relative
  concentration of the winning clone as a function of time.}
\end{figure}

\begin{figure}[p]
\centerline{
\includegraphics[width=.7\columnwidth]{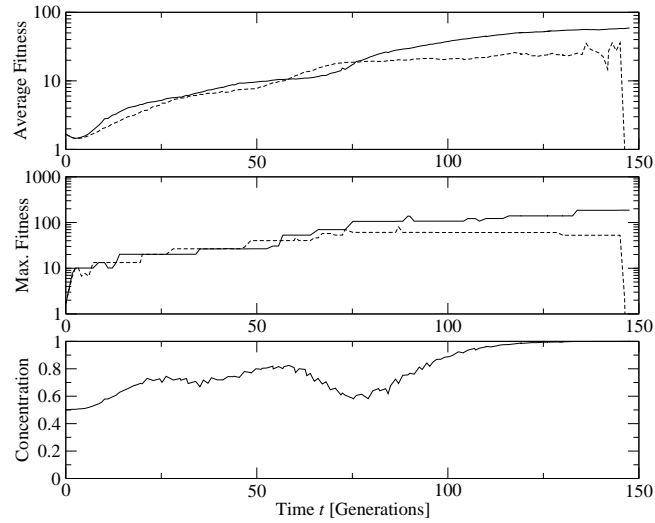}
}
\caption{\label{fig:compN10000-B}Typical dynamic of a competition experiment
  for strain B ($N=10,000$). The top two graphs show the average and the
  maximum fitness of the winning (solid lines) and the losing (dashed lines)
  clones as a function of time. The bottom graph shows the relative
  concentration of the winning clone as a function of time.}
\end{figure}

\begin{figure}[p]
\centerline{
\includegraphics[width=.7\columnwidth]{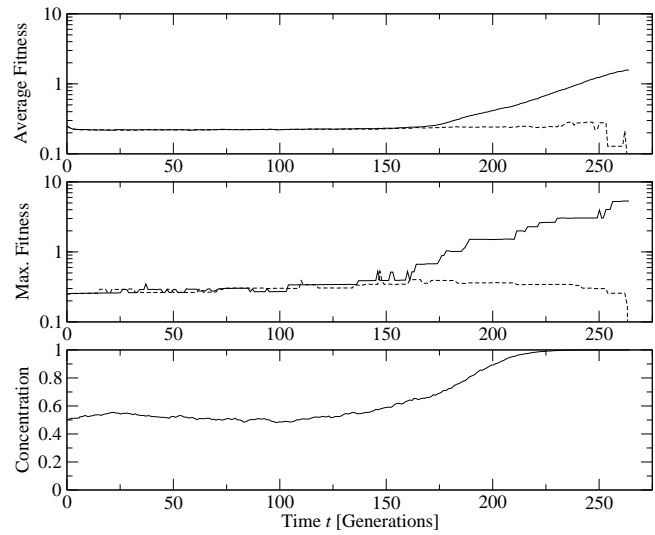}
}
\caption{\label{fig:compN10000-C}Typical dynamic of a competition experiment
  for strain C ($N=10,000$). The top two graphs show the average and the
  maximum fitness of the winning (solid lines) and the losing (dashed lines)
  clones as a function of time. The bottom graph shows the relative
  concentration of the winning clone as a function of time.}
\end{figure}

Typical competition dynamics for the three strains are displayed in
Figs.~\ref{fig:compN10000-A}-\ref{fig:compN10000-C}. In all three figures, I
show the average and maximum fitness of the winning and the losing clone as a
function of time, as well as the relative concentration of the winning clone
as a function of time. 

For strain A, average and maximum fitness of the two competing clones remained
typically almost unchanged until the losing clone disappeared
(Fig.~\ref{fig:compN10000-A}). Changes in relative clone size were mostly due
to genetic drift. However, the theoretical prediction
Eq.~\eqref{eq:extinct-neutral} works only moderately well for the smallest
population size, and overestimates the extinction times for the larger
population sizes substantially (Fig.~\ref{fig:extinct-times}). I verified that
this effect is not a shortcomming of the theoretical prediction by carrying
out additional competition experiments with all mutations switched off (data
not shown).  For these experiments, the extinction data were in very good
agreement with the theoretical prediction, which implies that the reduced
extinction times in the competitions with mutations must be due to the
accumulation of deleterious mutations. Indeed, in Fig.~\ref{fig:compN10000-A},
we see that the maximum fitness of the losing clone is---after approximately
500 generations---consistently lower than the maximum fitness of the winning
clone. Even though the difference in maximum fitness between the two
clones is very small, it is sufficient to accelerate the extinction process.
And the smaller the losing clone becomes, the more likely it is to experience
even further reductions in its maximum fitness. The final stage of the
competition is mutational
meltdown~\cite{LynchGabriel90}: Decreasing clone size accelerates loss of the
highest fitness mutants, which in turn results in even further reduction of
clone size. The clone decreases in size and loses fitness
at an ever accelerating pace, until it has disappeared.

Fig.~\ref{fig:compN10000-A} clearly shows that mutational meltdown takes place
towards the end of the competition and leads to a reduced extinction time.  At
first glance it is surprising that mutational meltdown should be responsible
for the increasing deviations between theory and measured extinction times as
the population size increases. After all, mutational meltdown is commonly
associated with small population sizes. The reason why this effect here
becomes more pronounced at larger population sizes is the following: When the
relative difference in fitness between two mutants is smaller than the inverse
of the population size, then these two mutants are effectively neutral, in the
sense that they are equally affected by genetic drift. Therefore, larger
population sizes can resolve finer fitness differences between mutants. In the
case of strain A, the fitness difference between the winning and the losing
clone is miniscule, so that at small population sizes, drift is the dominant
factor. Once the population size is sufficiently large, however, this small
fitness difference turns the population dynamic deterministic, and the clone
that loses the fittest mutant first will slowly but surely disappear.

A competition between two clones of strain C typically started out just like
one between two clones of strain A. However, often one of the two clones
managed eventually to acquire a beneficial mutation with substantial selective
advantage, and would then quickly exclude the other clone. Since beneficial
mutations were fairly rare for strain A, the second clone did not have the
chance to pick up an even better mutation in the short time that remained
before it died out. Clearly, the time to the first beneficial mutation
determined therefore the distribution of extinction times, unless it was much
larger than the typical time to extinction by genetic drift. The time to the
first beneficial mutation grows with decreasing population size, while the
time to extinction by drift grows with increasing population size. The
distributions of extinction times for strain C are determined by these two
constraints: For $N=100$, beneficial mutations are very rare, and the
extinction times are dominated by the effects of drift. For $N=1000$,
beneficial mutations are still rare, but nevertheless sufficiently abundant,
so that the extinction times are clearly shorter than the ones for drift
alone. Finally, for $N=10,000$, beneficial mutations are so frequent that the
time to extinction is on average even shorter than for $N=1000$.

Strain B showed a competition dynamic very similar to the one described by
Sol\'e et al.~\cite{Soleetal99}. Both the marked and the unmarked clone gained
substantially in fitness during the competition, and both clones would
alternatingly take the lead in fitness gains
(Fig.~\ref{fig:compN10000-C}). However, this Red-Queen dynamic came at a
price: The time to extinction of either clone was consistently shorter than
for strains A or C. Apparently, the constantly changing relative growth rates
of the two competing clones introduced increased fluctuations in the clone
sizes, so that one of the clones was quickly reduced to a size at which it
became prone to mutational meltdown, or was at least substantially impaired in
its ability to acquire further beneficial mutations.

\section{Conclusions}

The location in fitness space from where a strain is taken has a strong
influence on its competition dynamic. A clear arms race of mutants with ever
increasing fitness can only be observed when beneficial mutations are
abundant. When beneficial mutations are rare or completely absent, then either
the clone that finds a beneficial mutation first wins or the clone that loses
the highest-fitness mutant first loses. In general, it seems that a positive
mutation rate will always reduce the competition time, so that the loser dies
out earlier than it would in the absence of mutations. The Red-Queen dynamic,
where both clones acquire mutants of ever increasing fitness, is particularly
unstable. In this case, competitions last the shortest.

The results that I have presented here were obtained in computer experiments
with digital organisms. Therefore, it is not a priori clear that my
conclusions apply directly to RNA viruses. Nevertheless, I believe it is very
likely that they do. In particular, the way in which RNA virus strains are
typically prepared for competition experiments (frequent transfers at small
population sizes, which leads to accumulation of deleterious mutations, and
transfers to new environments, where many advantageous mutations can be
acquired) are similar to my preparation of strain B. The fact that they show
competition dynamics very similar to that of strain B is therefore reassuring.
To test my other predictions in a virus system, one would need strains
similar to A or C. In principle, it should be possible to
prepare a virus strain which is located at or near the top of a fitness peak,
by propagating the virus for a long time in a constant environment at a large
effective population size. With such a strain, the competition dynamic should
be more similar to my strain A, or maybe C, than B. If this turns out to be
true, then competition experiments at various population sizes can be used as
a reliable tool to map out the neighborhood in fitness space of a particular
virus strain.

\section*{Acknowledgments}
This work was supported by the NSF under contract No. DEB-9981397. I would
like to thank D.~A. Drummond for interesting discussions on coalescent theory,
and C. Adami for helpful comments on the manuscript.

\bibliographystyle{unsrt}
\bibliography{paper}
\end{document}